\documentclass[prb,twocolumn,superscriptaddress,showpacs,floatfix]{revtex4}

\usepackage{graphicx}
\usepackage{amsfonts}
\usepackage{amsmath}

\newcommand{\mystate}[3]{\left| { {#1} \atop {#2}} #3\right>}

\begin{document}

\title{Optical study of electron-electron exchange interaction in
CdTe/ZnTe quantum dots}

\author{T. Kazimierczuk}
\email{Tomasz.Kazimierczuk@fuw.edu.pl}
\affiliation{%
Institute of Experimental Physics, Faculty of Physics, University of
Warsaw,\\
Ho\.za 69, 00-681 Warsaw, Poland }%

\author{T. Smole\'nski}
\affiliation{%
Institute of Experimental Physics, Faculty of Physics, University of
Warsaw,\\
Ho\.za 69, 00-681 Warsaw, Poland }%

\author{J. Kobak}
\affiliation{%
Institute of Experimental Physics, Faculty of Physics, University of
Warsaw,\\
Ho\.za 69, 00-681 Warsaw, Poland }%

\author{M. Goryca}
\affiliation{%
Institute of Experimental Physics, Faculty of Physics, University of
Warsaw,\\
Ho\.za 69, 00-681 Warsaw, Poland }%

\author{W. Pacuski}
\affiliation{%
Institute of Experimental Physics, Faculty of Physics, University of
Warsaw,\\
Ho\.za 69, 00-681 Warsaw, Poland }%

\author{A. Golnik}
\affiliation{%
Institute of Experimental Physics, Faculty of Physics, University of
Warsaw,\\
Ho\.za 69, 00-681 Warsaw, Poland }%

\author{K. Fronc}
\affiliation{%
Institute of Physics, Polish Academy of Sciences,\\
Al. Lotnik\'ow 32/64, 02-688 Warsaw, Poland}%

\author{\L{}. K\l{}opotowski }
\affiliation{%
Institute of Physics, Polish Academy of Sciences,\\
Al. Lotnik\'ow 32/64, 02-688 Warsaw, Poland}%

\author{P. Wojnar}
\affiliation{%
Institute of Physics, Polish Academy of Sciences,\\
Al. Lotnik\'ow 32/64, 02-688 Warsaw, Poland}%

\author{P. Kossacki}
\affiliation{%
Institute of Experimental Physics, Faculty of Physics, University of
Warsaw,\\
Ho\.za 69, 00-681 Warsaw, Poland }%

\date{\today}

\begin{abstract}
We present an experimental study of electron-electron exchange interaction
in self-assembled CdTe/ZnTe quantum dots based on the photoluminescence
measurements.
The character and strength of this interaction are obtained by simultaneous
observation of various recombination channels of a doubly negatively charged
exciton X$^{2-}$, including previously unrecognized emission lines related
to the electron-singlet configuration in the final state. A typical value of the
electron singlet-triplet splitting, which corresponds to the exchange
integral of electron-electron interaction, has been determined as 20.4 meV
with a spread of 1.4 meV across the wide population of quantum dots. We
also evidence an unexpected decrease of energy difference between the singlet
and triplet states under a magnetic field in Faraday geometry.
\end{abstract}


\maketitle

\section{Introduction}
Semiconductor quantum dots (QDs) with their ability to confine carriers
create convenient conditions to study various interactions between
confined quasi-particles, in particular the exchange interaction.
Experimentally, polarization-resolved spectroscopy provides a relatively easy
access to the exchange interaction between an electron and a
hole.\cite{bayer-prb-2002} Due to heavy-hole character of the exciton
ground state in a QD, the effective electron-hole exchange interaction is
Ising-like (described by the so-called isotropic exchange constant $\delta_0$)
with a weak perturbation related to the QD in-plane anisotropy (described
by anisotropic exchange constants $\delta_1$ and $\delta_2$). 

In contrast to the electron-hole interaction, the exchange interaction
between two electrons is much closer to the picture of two simple
$1/2$-spin particles and thus leads to the formation of singlet and triplet
configurations. Such a singlet-triplet structure was mostly studied in the
coupled dot pairs \cite{johnson-prb-2005}. Observation of the effects
related to the singlet-triplet structure in a single self-assembled QD is
more difficult. Such effects manifest themselves only in the properties
of the excited excitonic complexes. Emergence of singlet-triplet structure
requires the presence of two electrons occupying different orbitals, e.g.,
one electron occupying the $s$-shell orbital, and the other one the
$p$-shell orbital.
The simplest excitonic complexes with such carriers are the excited
negatively charged exciton (${\mathrm{X}^-}^*$) and the doubly 
negatively charged exciton (X$^{2-}$) in the fundamental state.
The properties of electron-electron interaction can be accessed by
studying different recombination channels of an X$^{2-}$
complex\cite{urbaszek-prl-2003,finley-prb-2001,poem-prb-2007,ediger-prl-2007}
or the photoluminescence excitation (PLE) of negatively charged dots
\cite{akimov-pssa-2004}. The values of \textit{s}-\textit{p} electron-electron exchange
integral reported in previous studies vary from about 4 meV
\cite{finley-prb-2001} for InAs/GaAs QDs, to nearly 80 meV
\cite{akimov-pssa-2004} in the case of CdSe/ZnSe QDs.

In this paper we present a spectroscopic study of emission lines related
to different recombination channels of a doubly negatively charged exciton
X$^{2-}$ confined in self-assembled CdTe/ZnTe QD. The final states of
identified transitions correspond to the excited state of two electrons
forming either a singlet or a triplet configuration. We discuss the
anisotropy-related fine structure of these transitions as well as their
behavior in an external magnetic field, both in Faraday and Voigt
configurations. We demonstrate that the properties of observed transitions
can be understood within the frame of a simple spin Hamiltonian previously
used to describe X$^{2-}$ and XX$^{-}$ recombination to the triplet final
state \cite{kazimierczuk-prb-2011-exchange}. We also point out deficiencies
of such a simple model, namely incorrect predictions of the relative
intensities of optical transitions to the singlet final state and the
effects related to singlet-triplet splitting in the magnetic field.
Moreover, we propose an explanation of the former shortcoming as
originating from the presence of configuration mixing in a QD, 
the importance of which was already shown in several previous
studies.\cite{hawrylak_multix, tsmol_app_configuration_mixing}

\section{Samples and experimental setup}
The measurements were performed on samples containing a single layer of
self-organized CdTe QDs embedded in a ZnTe matrix. The samples were grown by
molecular beam epitaxy (MBE). The QD formation was induced by temporary
covering the CdTe formation layer with amorphous tellurium, as described in
Refs. \onlinecite{tinjod-apl-2003, wojnar-nanotechnology-2008}. The
spectroscopy of single quantum dots over a wide range of emission energies
was achieved by collecting the light through nanometer-size shadow mask
apertures on a sample surface. The apertures were manufactured by
depositing a nontransparent 100nm layer of gold onto a sample with
spin-casted 200~nm polystyrene beads, which were afterwards removed by
ultrasonic rinsing in trichloroethylene and methanol. 

\begin{figure*}[tb]
\includegraphics{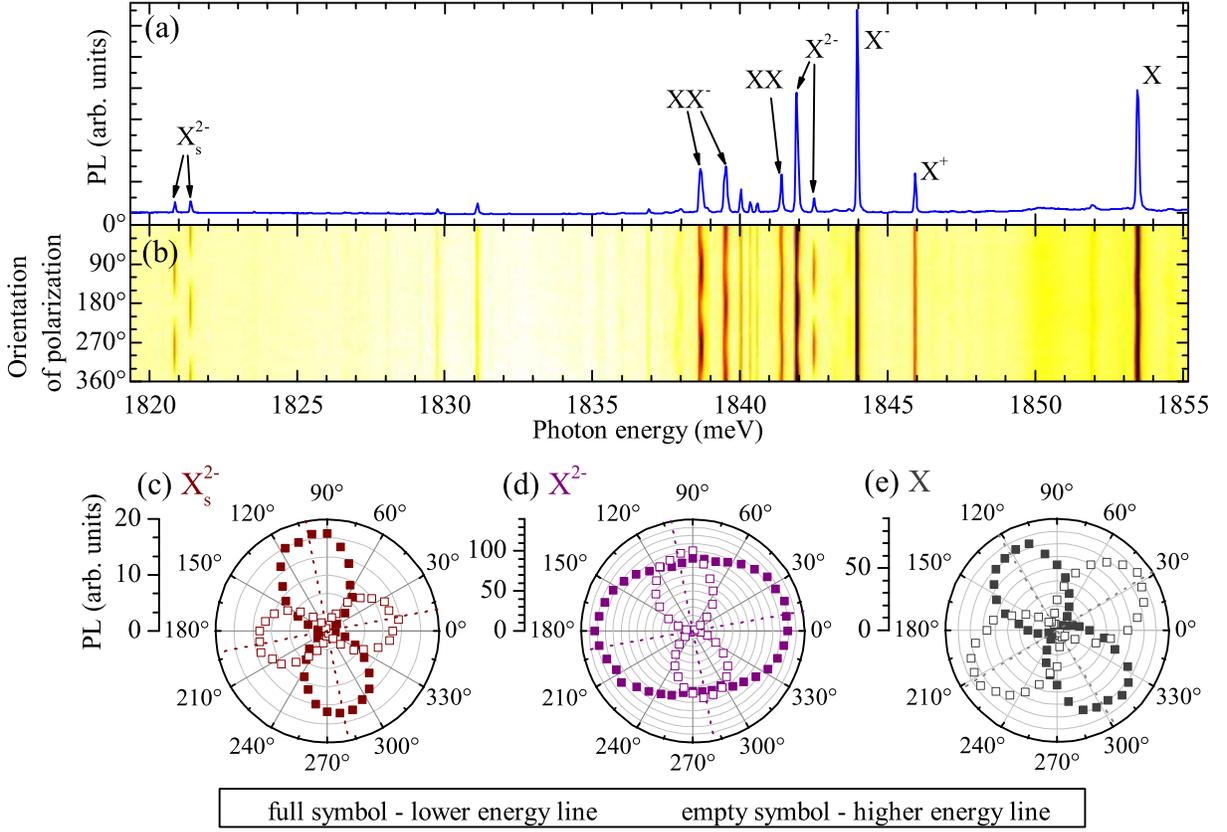}
\caption{(Color online) (a) Photoluminescence spectrum of a single
CdTe/ZnTe QD without polarization resolution. Lines denoted by
X$^{2-}_\mathrm{s}$ correspond to recombination of a doubly charged exciton
to a~state with two electrons in singlet configuration. 
(b) False-color plot presenting PL spectra of the same QD measured using
different linear polarization of detection.
The 0$^\circ$ and 90$^\circ$ correspond to cleavage axes of the sample.
(c-e) Dependence of the PL intensity of X$^{2-}_\textrm{s}$, X$^{2-}$, and
X doublets on the orientation of detected polarization for the same QD.
In case of X$^{2-}$, the lower energy line exhibit only partial linear
polarization due to the coincidental contribution from unpolarized
transition\cite{kazimierczuk-prb-2011-exchange}. The intensity of the higher
energy component of X$^{2-}$ was magnified by a factor 4 for better
visibility.
\label{fig:pl}}
\end{figure*}

The measurements were carried out in a high-resolution spectroscopy setup
with a laser spot of diameter corresponding to about 0.5~$\mu$m. The PL was
excited non-resonantly at 400 nm. The signal was resolved using a 0.5-m
monochromator and recorded using a CCD camera. The samples were placed in a
helium-bath cryostat at temperature of $1.5$ K. The cryostat was equipped
with a split-coil superconducting magnet producing magnetic field up to 11T
in Faraday or Voigt geometry. More details about the samples and the
experimental setup can be found in Refs. \onlinecite{kazimierczuk-prb-2010,
kazimierczuk-prb-2011-exchange}.

\section{Polarization-resolved photoluminescence of CdTe/ZnTe QDs}
Basic properties of measured single QDs PL spectra fully reproduce the
findings of previous studies of CdTe based
dots.\cite{kazimierczuk-prb-2011-exchange,suffczynski-prb-2006,besombes-prb-2002,klo11}
The PL spectrum of a single randomly selected QD [presented in Fig
\ref{fig:pl}(a)] consists of a~number of sharp emission lines originating
from recombination of different excitonic complexes. Due to the high
efficiency of single carrier capture under nonresonant
excitation\cite{suffczynski-prb-2006}, we observe emission lines related to
the transitions of excitons of various charge states. The sequence and the
relative energies of the emission lines were shown to be similar for
different self-assembled CdTe/ZnTe
QDs\cite{kazimierczuk-prb-2011-exchange}. On this basis we identify the
strongest emission lines in the spectrum presented in Fig. \ref{fig:pl}(a)
as related to the recombination of neutral exciton (X), positively and
negatively charged excitons (X$^+$, X$^-$), doubly negatively charged
exciton (X$^{2-}$), neutral biexciton (XX), and negatively charged
biexciton (XX$^-$). Most of these transitions exhibit non-trivial fine
structure due to the presence of exchange interaction between confined
carriers\cite{bayer-prb-2002,kazimierczuk-prb-2011-exchange}. Analysis of
this fine structure is a~potent tool in identification of the emission
lines in single dot PL spectrum. A well known example is distinguishing
between neutral and charged exciton\cite{bayer-prb-2002}. Here we use the
analysis of the fine structure to argue that despite the surprisingly large
spectral distance, a~previously unrecognized pair of lines denoted in Fig.
\ref{fig:pl} as X$^{2-}_\mathrm{s}$ does in fact originate from the same QD
and is related to the recombination of doubly charged exciton 
to the two-electron singlet state. 

The analysis of the fine structure benefits from the fact
that in the typical case, the exchange interaction between electrons and holes has
anisotropic character. This anisotropy is manifested by the linear
polarization of the light emitted during recombination of
fine-structure-split excitonic complexes. Previous studies revealed that
different emission lines of the same quantum dot can exhibit different
polarization axes\cite{kazimierczuk-prb-2011-exchange,poem-prb-2007},
especially when they originate from recombination of excitonic complexes
consisting of carriers occupying the higher shells. 
As a consequence, the polarization-resolved experiment cannot be reduced
to measurements of PL signal in two orthogonal linear polarizations.
Instead, the intensity of each emission line needs to be analyzed for a
number of different polarizations in order to determine the orientation of
polarization related to a given transition. Figure \ref{fig:pl}(b) shows PL
spectra of a single QD measured for a whole range of orientations of
detected linear polarization. Due to relatively small splitting, the fine
structure of the neutral exciton is visible here only as a small modulation
of the average energy of X (and XX) emission line. Nevertheless, the intensities of
both fine-split components of X could be extracted by fitting the emission
line with two Gaussian profiles of fixed positions. In the same way we also
extracted the angle-dependent intensities of other emission lines,
including X$^{2-}_\mathrm{s}$ lines as well. Angular dependence of the
intensity of the relevant lines is presented in Figs. \ref{fig:pl}(c)-\ref{fig:pl}(e). As
expected, in each case the anisotropic electron-hole exchange interaction
led to splitting of the emission line into two components of orthogonal
linear polarizations. Possible deviation from the orthogonality of these
linear polarizations is introduced by the valence band
mixing\cite{leger-prb-2007, koudinov-prb-2004, tsmol-dark-prb}, which for
the QD in Fig. \ref{fig:pl} was found to be smaller than 3\%. 

The QD shown in Fig. \ref{fig:pl} exhibits a difference between
orientations of the polarization of the neutral exciton X (here,
32$^\circ$) and the doubly charged exciton X$^{2-}$ (here: 11$^\circ$).
Interestingly, the orientation of the latter one exactly (with precision
better than 1$^\circ$) matches the fine structure of the previously
unrecognized X$^{2-}_\mathrm{s}$ doublet. Moreover both complexes ---
X$^{2-}$ and X$^{2-}_\mathrm{s}$ --- exhibit practically equal fine
structure splittings of $0.53$ meV. Such a coincidence strongly suggests
existence of a link between X$^{2-}$ and X$^{2-}_\mathrm{s}$. These
observations are not specific only to the presented QD.
In the following section we argue that these properties are characteristic
for the recombination of a doubly charged exciton X$^{2-}$ to the state
with two electrons in a singlet configuration (as opposed to 
the previously recognized recombination to triplet configuration) and compare the
predictions of the model calculations with the measurements on a population
of 18 different QDs.

\section{Predictions regarding  X$^{2-}$ transition within spin
Hamiltonian model}

\begin{figure}
\centering
\includegraphics{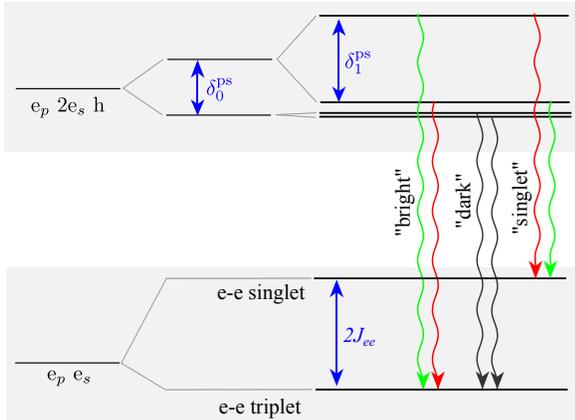}
\caption{(Color online) A scheme of states and transitions related to
the recombination of doubly charged exciton X$^{2-}$. Note that the level
spacing is not in scale. Colors of the arrows denote polarization of the
transition. Red and green (different shades of light grey) denote orthogonal
linear polarization while dark grey denotes unpolarized
transition.\label{fig:schemat}}
\end{figure}

The simplest and the most transparent way to model the recombination of
doubly charged exciton is to use simple spin Hamiltonian.
The model assumes that the spin-related interactions between confined
carriers are weak in comparison with the spatial confinement.
In such a case, the carriers populate subsequent orbitals according to the
aufbau principle, and the exchange interaction is used only to determine
the fine structure within otherwise degenerate subspace. 

The description of doubly charged exciton X$^{2-}$ required using three types of
carriers: \textit{s}-shell hole, \textit{s}-shell electron, and
\textit{p}-shell electron. For simplicity, we assume the presence of only
one lowest \textit{p} orbital, possibly due to the anisotropy of the dot.
Each pair of the carriers interact by the exchange interaction. The
interaction between \textit{s}- or \textit{p}-shell electron with the hole
includes both iso- and anisotropic parts
\cite{urbaszek-prl-2003,akimov-prb-2005,ediger-prl-2007,kazimierczuk-prb-2011-exchange}:
\begin{equation}
H_1 = 2 \delta_0 S^{(s)}_z \sigma_z + \delta_1  \left(S^{(s)}_x \sigma_x -
S^{(s)}_y \sigma_y \right),
\end{equation}
\begin{equation}
H_2 = 2 \delta^\mathrm{(ps)}_0 S^{(p)}_z \sigma_z + \delta^\mathrm{(ps)}_1
 \left(S^{(p)}_x \sigma_x - S^{(p)}_y \sigma_y \right),
\end{equation}
where $S^{(s)}$, $S^{(p)}$, and $\sigma$ are
the spin operators  for
\textit{s}-shell electron, \textit{p}-shell electron, and the hole
respectively.
In case of the hole, we use pseudo-spin convention, i.e., we formally
treat the $\pm 3/2$ heavy hole states as $\mp 1/2$ pseudo-spin states. The
inter-shell electron-electron exchange interaction is assumed to have
purely isotropic character, as the experimental data do not indicate such
an anisotropy:
\begin{equation}
H_3 = -2 J_\mathrm{ee} \vec{S}^{(s)}\vec{S}^{(p)}.
\end{equation}
The final Hamiltonian is a sum of such contributions for each pair of
confined carriers. In the presence of 
an external magnetic field, additional
terms of the form $g\mu_B B S$ 
are added for each confined carrier. 

The initial state of transition consists of three electrons and one hole.
Two electrons fully occupy the \emph{s}-shell and the remaining electron is
placed on the \emph{p}-shell.
As a result, there are four base spin configurations of an initial excitonic
complex with two possible spin orientations of the hole as well as the
\textit{p}-shell electron. The degeneracy of these four states is lifted by
the exchange interaction between the unpaired carriers. In this respect, there
is a close analogy between a neutral exciton (with a single electron on the
\textit{s} shell) and doubly charged exciton (with a single electron on the
\textit{p} shell). Consequently, the eigenstates of doubly charged exciton
take the form
\begin{center}
$\frac{1}{\sqrt{2}}\left(\mystate{\uparrow}{\uparrow\downarrow}{\Downarrow}
+ \mystate{\downarrow}{\uparrow\downarrow}{\Uparrow}\right)$,
$\frac{1}{\sqrt{2}}\left(\mystate{\uparrow}{\uparrow\downarrow}{\Downarrow}
- \mystate{\downarrow}{\uparrow\downarrow}{\Uparrow}\right)$,
$\frac{1}{\sqrt{2}}\left(\mystate{\uparrow}{\uparrow\downarrow}{\Uparrow}
+ \mystate{\downarrow}{\uparrow\downarrow}{\Downarrow}\right)$,
$\frac{1}{\sqrt{2}}\left(\mystate{\uparrow}{\uparrow\downarrow}{\Uparrow}
- \mystate{\downarrow}{\uparrow\downarrow}{\Downarrow}\right)$,
\end{center}
where $\mystate{A}{B}{C}$ denotes a state with $A$ electron(s) on the
\textit{p} shell, $B$ electron(s) on the \textit{s} shell and $C$ hole(s)
on the \textit{s} shell. $\uparrow$ and $\downarrow$ represent $\pm 1/2$
electron spin, while $\Uparrow$ and $\Downarrow$ represent $\pm 3/2$
heavy-hole spin.  The first and the second of the four states are the
analogs of the bright states of the neutral exciton. They are split by
anisotropic inter-shell exchange constant $\delta_1^\mathrm{ps}\approx
-0.5$ meV \cite{kazimierczuk-prb-2011-exchange}.  The third and the fourth
of these states are analogs of the dark states of the neutral exciton. The
splitting between two ``dark" states can be neglected in our dots. Both
pairs, ``bright" and ``dark" states are split by isotropic inter-shell
exchange constant $\delta_0^\mathrm{ps} \approx 0.3$ meV
\cite{kazimierczuk-prb-2011-exchange}.

In the final state of the transitions two electrons are left in the QD:
one on the \textit{s} shell and one on the \textit{p} shell. These two
electrons interact with
the exchange interaction, thus the eigenstates take a
form of singlet-triplet structure:
\begin{center}
$\left|S\right>=\frac{1}{\sqrt{2}}\left(\mystate{\downarrow}{\uparrow}{}-
\mystate{\uparrow}{\downarrow}{}\right)$,\\
$\left|T_1\right>=\mystate{\uparrow}{\uparrow}{}$, 
$\left|T_0\right>=\frac{1}{\sqrt{2}}\left(\mystate{\downarrow}{\uparrow}{}+
\mystate{\uparrow}{\downarrow}{}\right)$,
$\left|T_{-1}\right>=\mystate{\downarrow}{\downarrow}{}$.
\end{center}

Optical selection rules require antiparallel spin orientation of
recombining electron-hole pair. Moreover, due to symmetry difference
between \textit{s}- and \textit{p}-shell, only \textit{s}-shell electrons
can recombine with the \textit{s}-shell hole. Even if this restriction is
partially lifted by imperfections of the real QD potential, the assumed
recombination of \textit{s}-shell electron should be  significantly
stronger than recombination involving \textit{p}-shell electron. More
importantly, recombination of \textit{p}-shell electron would produce
emission lines at higher energy than other emission lines in the studied PL
spectrum, which we do not find in the experiment. 

Discussed selection rules constraints are met by three sets of
transitions.
The first group of transitions is related to initial states being the
analogs of the dark states of a neutral excitons (i.e., the unpaired
electron and the hole have parallel spin orientation). Consequently we will
refer to these transitions using label ``dark", irrespectively of the
actual strength of these 
transitions. According to the selection rules, the 
\textit{s}-shell electron remaining after recombination has the same
orientation of the spin as the recombining hole, which for ``dark" state is
the same as the \textit{p}-shell electron. Consequently, the allowed final
states of the transition are $S=\pm 1$ states of the triplet with two
parallel electron spins:
\begin{center}
$\frac{1}{\sqrt{2}}\left(\mystate{\uparrow}{\uparrow\downarrow}{\Uparrow}
+ \mystate{\downarrow}{\uparrow\downarrow}{\Downarrow}\right)
\xrightarrow{\mathrm{``dark"}} 
\mystate{\uparrow}{\uparrow}{}  \mathrm{or} 
\mystate{\downarrow}{\downarrow}{}$, \\
$\frac{1}{\sqrt{2}}\left(\mystate{\uparrow}{\uparrow\downarrow}{\Uparrow}
- \mystate{\downarrow}{\uparrow\downarrow}{\Downarrow}\right)
\xrightarrow{\mathrm{``dark"}} 
\mystate{\uparrow}{\uparrow}{}  \mathrm{or} 
\mystate{\downarrow}{\downarrow}{}$.
\end{center}
Without magnetic field neither initial nor the final state of these
transitions are split. As a result, these transitions contribute to a
single non-polarized line in the PL spectrum.

The second and the third group of the transitions are related to
recombination of the X$^{2-}$ complex with anti-parallel spins of the
\textit{p}-shell electron and the hole. According to the selection rules,
the electrons in the final state of the transition also have anti-parralel
spin configuration which may indicate either the singlet state or the $S=0$
triplet state. A more strict approach\cite{kazimierczuk-prb-2011-exchange}
shows, that our model entails both possible final states. We will refer to
the transitions leading to the triplet state using the label ``bright" (again,
with no reference to the transition strength), while the transitions
leading to the singlet state will be referred 
to simply as ``singlet":
\begin{center}
$\frac{1}{\sqrt{2}}\left(\mystate{\uparrow}{\uparrow\downarrow}{\Downarrow}
+ \mystate{\downarrow}{\uparrow\downarrow}{\Uparrow}\right)
\xrightarrow{\mathrm{``bright"}}
\frac{1}{\sqrt{2}}\left(\mystate{\uparrow}{\downarrow}{} +
\mystate{\downarrow}{\uparrow}{}\right)$, \\
$\frac{1}{\sqrt{2}}\left(\mystate{\uparrow}{\uparrow\downarrow}{\Downarrow}
- \mystate{\downarrow}{\uparrow\downarrow}{\Uparrow}\right)
\xrightarrow{\mathrm{``bright"}}
\frac{1}{\sqrt{2}}\left(\mystate{\uparrow}{\downarrow}{} +
\mystate{\downarrow}{\uparrow}{}\right)$, \\
$\frac{1}{\sqrt{2}}\left(\mystate{\uparrow}{\uparrow\downarrow}{\Downarrow}
+ \mystate{\downarrow}{\uparrow\downarrow}{\Uparrow}\right)
\xrightarrow{\mathrm{``singlet"}}
\frac{1}{\sqrt{2}}\left(\mystate{\uparrow}{\downarrow}{} -
\mystate{\downarrow}{\uparrow}{}\right)$, \\
$\frac{1}{\sqrt{2}}\left(\mystate{\uparrow}{\uparrow\downarrow}{\Downarrow}
- \mystate{\downarrow}{\uparrow\downarrow}{\Uparrow}\right)
\xrightarrow{\mathrm{``singlet"}}
\frac{1}{\sqrt{2}}\left(\mystate{\uparrow}{\downarrow}{} -
\mystate{\downarrow}{\uparrow}{}\right)$.
\end{center}

All these transitions are depicted in Fig. \ref{fig:schemat}. The spin
Hamiltonian model predicts that ``singlet" transitions should be present in
the PL spectrum as a pair of lines with exactly the same splitting as
``bright" transitions ($\delta_1^{ps}$). In both cases the split lines
exhibit orthogonal linear polarizations. The orientation of this linear
polarization is governed by the (common) initial state of the 
transition
and therefore should be the same for ``singlet" and ``bright" transitions.
However, due to different symmetry of the final states, the order of the
polarizations is reversed. The lower energy line of ``singlet" transition has
the same polarization as higher energy line of ``bright" transition and
\textit{vice versa}.

\begin{figure}
\includegraphics{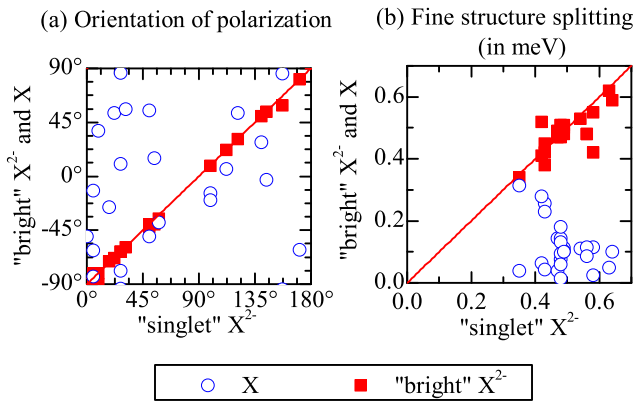}
\caption{(Color online) Correlation of (a) orientations of polarization
and (b) splitings of doubly charged exciton 
X$^{2-}$ for different QDs (red
squares). Blue circles represent analog data for the neutral exciton for
comparison. Solid line corresponds to 
the predictions of the spin Hamiltonian
model, namely 90$^\circ$ difference in orientation of polarization and the
same splitting of ``bright" and ``singlet" transitions. \label{fig:stats}}
\end{figure}

These predictions are fully confirmed by the experimental data, as shown
in Fig. \ref{fig:stats}. Despite random distribution of the polarization
angle or the value of the splitting, for each studied QD ``singlet"
transition exhibited the same splitting and opposite linear polarizations
as ``bright" recombination of X$^{2-}$. Such a good agreement is an
unequivocal confirmation of correct identification of line 
X$^{2-}_\mathrm{s}$
from Fig.
\ref{fig:pl} as a ``singlet" recombination channel for doubly charged
exciton X$^{2-}$.

\begin{figure}
\centering
\includegraphics[width=7.93cm]{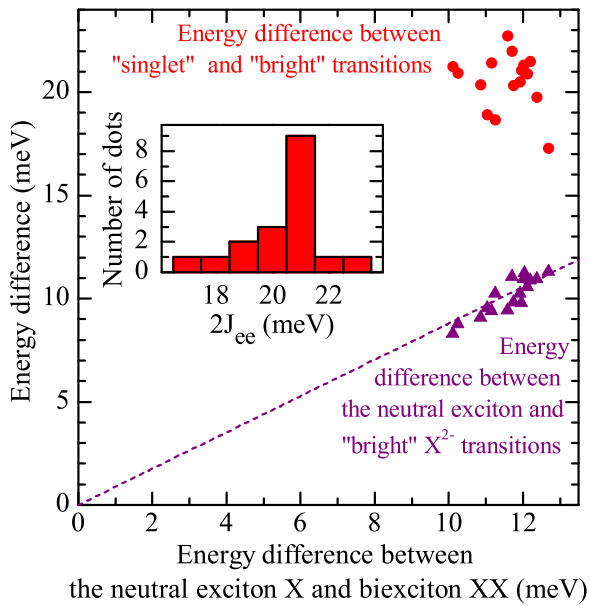}
\caption{(Color online) Correlation of relative spectral line positions. The relative
position of ``bright" and ``dark" transitions is proportional to the
relative position of the other lines of the same QD (e.g., neutral
biexciton), while the splitting between singlet and triplet
($2J_\mathrm{ee}$) seems to be
independent on these relative positions. The inset
presents a histogram of $2J_\mathrm{ee}$ constant among studied dots. 
\label{fig:scaling}}
\end{figure}

Simultaneous observation of both ``singlet" and ``bright" transition lines
opens a possibility to determine a strength of electron-electron exchange
constant $J_\mathrm{ee}$. The value determined by averaging a~distance
between ``singlet" and ``bright" emission lines for 18 different QDs
yielded $2J_\mathrm{ee} = \left(20.4 \pm 1.4\right)$ meV. Interestingly,
this quantity does not scale with energy distances between other lines,
which otherwise vary proportionally from dot to
dot\cite{kazimierczuk-prb-2011-exchange} in our samples. It is shown in
Fig. \ref{fig:scaling}, where clear proportionality between energy
distances from X to XX and from X to ``bright" transition of X$^{2-}$
contrasts with a lack of correlation for $J_\mathrm{ee}$ value.

We also stress out, that our measurements do not indicate the presence of
the anisotropic part of electron-electron exchange interaction, which was
anticipated in Ref. \onlinecite{ware-prb-2005}. Such an anisotropy of the
exchange interaction would
lift the degeneracy between $T_0$ and $T_{\pm 1}$ states. In such a
scenario the higher-energy ``bright" line would be a transition from
non-degenerate initial state to non-degenerate final state and thus
cannot exhibit any splitting. However, the measurements reported earlier in
Ref. \onlinecite{kazimierczuk-prb-2011-exchange} clearly show two-fold
linear splitting of this emission line in the magnetic field in the Voigt
geometry. Observed behavior is a~clear proof of degeneracy of $T_0$ and
$T_{\pm 1}$ states at $B=0$, which indicates purely isotropic character of
the electron-electron exchange interaction.

\section{Doubly charged exciton in magnetic field}
\begin{figure}
\includegraphics{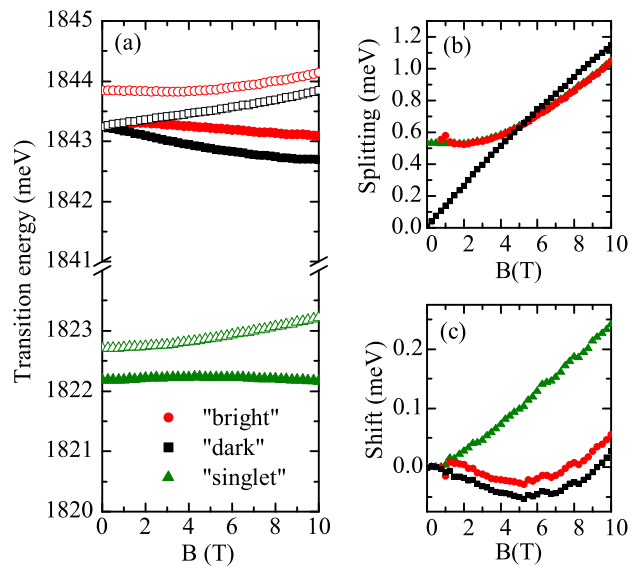}
\caption{(Color online) (a) Magnetic field dependence of transition
energies in Faraday configuration. Empty (full) symbols denote lines with
$\sigma+$ ($\sigma-$) helicity. The same data was used to calculate (b)
Zeeman splitting and (c) diamagnetic shift for each pair of lines.
\label{fig:faraday}}
\end{figure}

In order to provide additional information about studied transitions, we
measured the magnetic field dependence of all transitions related to
recombination of doubly charged exciton. Results of such an experiment in
Faraday configuration are shown in Fig. \ref{fig:faraday}. 

Under magnetic field along the growth axis, we observed clear splitting of
each set of transitions (``bright", ``dark", and ``singlet") into two lines
of opposite helicity of polarization. For ``bright" and ``singlet"
transitions, the degree of circular polarization was increasing with field
intensity, while ``dark" transitions exhibited full circular polarization
even in moderate magnetic field. Indeed, according to our model, magnetic
field in Faraday configuration does not couple ``bright" and ``dark"
initial states nor singlet-triplet final states. As a result, effect of the
magnetic field can be analyzed separately for each pair of X$^{2-}$
transitions lines. 

We first focus on the splitting of each pair of lines [see Fig.
\ref{fig:faraday}(b)]. Two main contributions to this splitting are the
zero-field fine structure splitting and the Zeeman contribution. As
discussed earlier, the fine structure splitting is present for both
``bright" and ``singlet" lines, while ``dark" transitions do not exhibit
observable zero-field splitting. Under influence of the field, the
splitting of each line increases according to the relation $\sqrt{\Delta^2
+ \left(g\mu_BB\right)^2}$, where $\Delta$ is
corresponding zero-field splitting.
As reported previously\cite{kazimierczuk-appa-2011}, we find different
effective $g$-factor values for the ``bright" and ``dark" transitions,
which we interpret as a signature of difference in $g$-factors for
\textit{s}- and \textit{p}-shell 
electrons. In comparison, we find the field dependence of the splitting of the
``singlet" lines to perfectly follow the dependence of the splitting of the
``bright" lines. This is another confirmation of the fact, that both of
these pairs of transitions originate from the same initial states and lead to a
single (but different in each case) final state.

\begin{figure}
\includegraphics{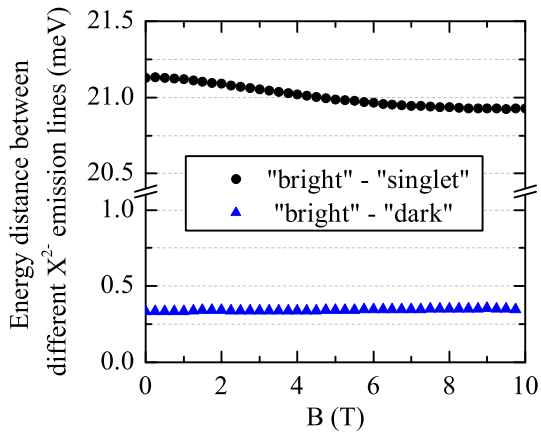}
\caption{(Color online) Energy distances between various pairs of emission
lines related to X$^{2-}$ recombination in magnetic field in Faraday
geometry. Energy of each pair is defined by the average of both components
using data from Fig. \ref{fig:faraday}(a). Due to common initial state,
''bright"-''singlet" distance is a direct measure of splitting between
singlet and triplet $T_0$ configuration of two electrons.  
\label{fig:jee_faraday}}
\end{figure}

\section{Beyond the spin Hamiltonian model}

This close connection between the ``singlet" and ``bright" transitions allows us to
directly measure the electron singlet-triplet energy distance as a function
of the magnetic field. In principle, neither the electron singlet nor the
triplet $T_0$ state should be affected by the magnetic field. However, the
experimental results shown in Fig. \ref{fig:jee_faraday} indicate that
magnetic field leads to a small reduction in the distance between the
energy of electron singlet state and triplet $T_0$ state. The presence of
this effect was confirmed by repeating the experiment on several other QDs.
The physical origin of the observed variation is not fully understood.
Possibly, the effect is due to different diamagnetic shift of the singlet
and triplet state. Increased spatial extension of the electron singlet
state due to additional exchange energy would lead to greater diamagnetic
shift and thus would reduce the energy difference between the singlet and
triplet state.
However, the range of magnetic field available in our experiments do not
allow us to unequivocally verify this hypothesis.

\begin{figure}
 \includegraphics{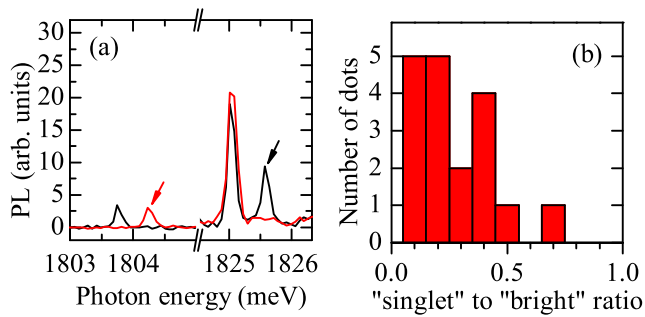}
 \caption{(Color online) (a) Photoluminescence spectrum with emission
lines related to different recombination channels of X$^{2-}$ measured
in two orthogonal linear polarizations. Arrows point the transitions
originating from the same initial state either to singlet or triplet
configuration. The intensities of these two channels should be equal
according to the spin hamiltonian model. The actual intensity of the
``singlet" transition for this dot is only $0.45$ of the intensity of the
``bright" transition.
(b) Histogram of ratio between intensities of "singlet" and "bright"
emission lines for different QDs. The average
ratio for the studied population was $0.28$.
 \label{fig:intensity}}
 \end{figure}

The second notable difference between the predictions of the simple spin
Hamiltonian model and the experimental data is related to relative
intensities of different X$^{2-}$ emission lines. In particular, the
``singlet" and ``bright" transitions originate from the same initial states
and therefore their relative intensities do not depend on the QD excitation
but they are governed solely by the relative strength of dipole moment for
these transitions. The spin Hamiltonian model assumes that the spatial part
of the wavefunction is the same in all engaged states and thus the
transition moments are defined by the projection of the spin part of the
wavefunction onto possible final eigenstates. As such, the model predicts
equal strength of ``singlet" and ``bright" transitions. On the contrary,
the experimental data evidence a~significant systematic difference between
intensities of the emission lines related to ``singlet" and ``bright"
transitions, as shown in Fig. \ref{fig:intensity}. In order to understand
the possible origins of such a difference, we performed calculations of
oscillator strengths of X$^{2-}$ optical transitions within the frame of an
extended theoretical model, which takes into account the configuration
mixing \cite{hawrylak_multix, tsmol_app_configuration_mixing}. In contrast
to the previously exploited simple spin Hamiltonian model, the current one
includes the mixing between the many-partcile states consisting of the same
number of electrons and holes, which occupy different QD shells. Therefore,
in the extended model the spatial parts of the wavefunctions of X$^{2-}$
recombination final states can in principle be different, and thus the
predicted strengths of ``singlet" and ``bright" transitions may not be
equal. Such an analysis is carried out by detailed numerical calculations
performed in the basis of many-particle states involving three lowest QD
shells (i.e., up to $d$-shell), for typical CdTe/ZnTe QD parameters
described in Ref. \onlinecite{tsmol_app_configuration_mixing}. As a result
we obtained the ratio of ``singlet'' and ``bright'' transitions strength to
be about 0.5, which is much closer to the experimentally
obtained values (see Fig. \ref{fig:intensity}). However, the ratio given by the
extended model still does not reach the measured values, which indicates the
need for further investigation of that effect.

\section{Conclusions}
In conclusion, we characterized the interaction between two electrons in
CdTe/ZnTe QDs based on the results of the single dot PL measurements.
The two electron states were accessed by analysis of emission lines
related to three different recombination channels of doubly charged exciton
X$^{2-}$. Two of the channels (labeled as ``bright" and ``dark") lead to
the triplet configuration of remaining electrons, and the third one
(labeled as ``singlet") leads to the singlet configuration. By
identification of all these transitions in a~single dot~PL~spectrum we
evaluated the basic quantity describing electron-electron exchange
interaction in our CdTe/ZnTe QDs, namely the exchange constant
$2J_\mathrm{ee} = \left(20.4 \pm 1.4 \right)$ meV. We did not find any
fingerprints of anisotropic contribution to
the electron-electron interaction.
Moreover, we found that the energy splitting between the singlet and the
triplet state is reduced by external magnetic field parallel to the growth
axis. The origin of this effect is not fully understood yet. Some light on this
origin might be shed by the quantititative modeling of CdTe/ZnTe QDs,
which in turn can benefit from our findings on electron-electron exchange
interaction in this system.

\begin{acknowledgments}
This work was supported by the Polish Ministry of Science and Higher
Education in years 2012--2016 as research grants ``Iuventus" and a
``Diamentowy Grant", NCBiR project LIDER, and NCN projects DEC-2011/
01/B/ST3/02406 and DEC-2011/02/A/ST3/00131. Experiments were carried out
with the use of CePT, CeZaMat and NLTK infrastructures financed by the
European Union - the European Regional Development Fund within the
Operational Programme ``Innovative economy" for 2007--2013.
\end{acknowledgments}


\end{document}